\documentclass[manuscript,screen,nonacm]{acmart}

\usepackage{booktabs} 
\usepackage{comment}
\usepackage{graphicx}

\newenvironment{ParticipantQuote}%
{\begin{small}\begin{list}{0}{%
\setlength{\topsep}{0.0pt}%
\setlength{\itemsep}{0.6pt}%
\setlength{\leftmargin}{0.2in}%
\setlength{\itemindent}{0.0in}%
\setlength{\rightmargin}{0.1in}%
\setlength{\labelsep}{0.25em}}}%
{\end{list}\end{small}}

{\begin{list}{0}{\setlength{\leftmargin}{0.5in}\setlength{\itemindent}{-0.4in}\setlength{\labelsep}{0.25em}\setlength{\itemsep}{3pt}}}%
{\end{list}}

{\begin{list}{$\bullet$}{\setlength{\leftmargin}{0.05in}\setlength{\labelsep}{0.5em}}}%
{\end{list}}

\AtBeginDocument{%
  \providecommand\BibTeX{{%
    \normalfont B\kern-0.5em{\scshape i\kern-0.25em b}\kern-0.8em\TeX}}}

\setcopyright{acmlicensed}
\copyrightyear{2018}
\acmYear{2018}
\acmDOI{XXXXXXX.XXXXXXX}

\acmConference[Conference acronym 'XX]{Make sure to enter the correct
  conference title from your rights confirmation emai}{June 03--05,
  2018}{Woodstock, NY}
\acmISBN{978-1-4503-XXXX-X/18/06}

\begin{document}

\title{Supporting Annotators with Affordances for Efficiently Labeling Conversational Data}


\author{Austin Z. Henley}  
\affiliation{
  \institution{University of Tennessee}
  \country{USA}
}

\author{David Piorkowski}
\affiliation{
    \institution{IBM Research}
    \country{USA}
}
\email{djp@ibm.com}

\renewcommand{\shortauthors}{Henley and Piorkowski}

\begin{abstract}
Without well-labeled ground truth data, machine learning-based systems would not be as ubiquitous as they are today, but these systems rely on substantial amounts of correctly labeled data. Unfortunately, crowdsourced labeling is time consuming and expensive. To address the concerns of effort and tedium, we designed CAL, a novel interface to aid in data labeling. We made several key design decisions for CAL, which include preventing inapt labels from being selected, guiding users in selecting an appropriate label when they need assistance, incorporating labeling documentation into the interface, and providing an efficient means to view previous labels. We implemented a production-quality implementation of CAL and report a user-study evaluation that compares CAL to a standard spreadsheet. Key findings of our study include users using CAL reported lower cognitive load, did not increase task time, users rated CAL to be easier to use, and users preferred CAL over the spreadsheet.
\end{abstract}





\maketitle

\section{Introduction}
\label{sec-introduction}

Without well-labeled ground truth data, today's machine-learning-based systems would not nearly be as ubiquitous as they are today.
These systems rely to substantial amounts of correctly labeled data. 
In some cases, this data is relatively straightforward to label correctly, as is the case of using data from clinical trials' or log data from users' interactions. 
However, for some labeling tasks, such as those that are qualitative instead of quantitative, human judgment is required. 
Due to the relatively large amounts of data required for ground truth, the most common approach for labeling is to crowdsource such tasks.

Unfortunately, crowdsourced labeling is fraught with difficulties.
The fundamental problem with labeling is that it often produces noisy data, as labeling is colored by the perspectives of the individual annotator who may label differently than his or her peers. 
It is also time-consuming, expensive, and error-prone, especially in cases where annotators label without specialized tools and instead use commonly-available software.
In particular, it has been reported that annotators often use spreadsheets~\cite{Renner2003}, which we also found from our informal surveys at a large technology company.
Consequently, reducing the amount of labeled data necessary via new machine learning techniques has been a popular research topic for several years~\cite{arora2009estimating,blum1997ai,culotta2005AAAI,fu2013survey,settles2008EMNLP,wang2011active}, but even these approaches require some amount of consistently labeled data in order to be effective. 

Researchers have already begun developing formative methods and tools to improve labeling efficiency and quality. 
One notable tool is Aeonium, a visual-analytics interface designed to support qualitative labeling tasks through affordances such as visual overviews and keyword distributions to support multiple coders working on the same task ~\cite{Drouhard2017PacificVis}.
Another tool is Revolt, which leverages disagreements between annotators to improved label quality ~\cite{chang2017revolt}. 

Despite these advancements in tools for supporting labeling tasks, many barriers regarding the usability of such tools still remain.

To address the concern of effort and tedium, we designed CAL, a novel interface to aid annotators in efficiently labeling conversational data.
We made several key design decisions for CAL focusing on reducing errors that annotators make and providing easy access to information to help improve annotator recall.
Affordances that we developed in CAL include preventing inappropriate labels from being selected, in-situ guidance for selecting a label, incorporation of labeling guidelines into the interface, and access to prior labels for similar examples. 
These key design decisions were selected to address the established challenges annotators face when labeling large amounts of data.

In this paper, we present a production quality implementation of CAL and report a user-study evaluation that compares annotators labeling using CAL to those using a commonly used alternative, a spreadsheet. 
In particular, our user study addressed the following research questions: 
\begin{itemize}
\setlength\itemsep{0em}

\item
RQ1: Did CAL users have lower cognitive load during labeling tasks than spreadsheet users?

\item
RQ2: Do CAL users complete labeling tasks in less time than spreadsheet users?

\item 
RQ3: Did CAL users rate CAL to be easier to use than the spreadsheet?

\item
RQ4: Did users prefer using CAL over the spreadsheet?

\end{itemize}

\section{Background \& Related Work}
\label{sec-background}

\begin{figure}
\centering
\includegraphics[scale=1.1]{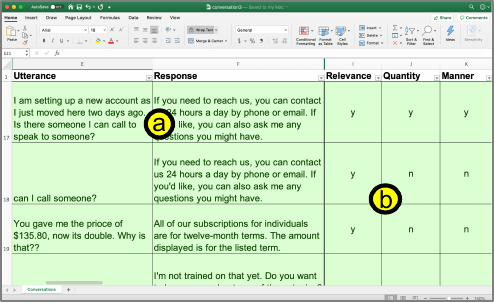}

\vspace{0pt}
\caption{%
An example spreadsheet used for labeling conversational data.
It includes the (a) transcript with the human utterance in the first column and the chatbot's response in the second column.
The annotator enters \emph{y} or \emph{n} into the (b) three columns representing different categories in this example code set.
}
\label{fig-spreadsheet}
\end{figure}

\subsection{Data Labeling}

Machine learning-based systems require large sets of data to have \emph{labels} that form the ground truth of supervised machine learning models. (Other similar terms include annotations, tags, codes, and classifications.)
To produce these labels, human \emph{annotators} manually look at the data, consisting of many \emph{examples} and select an appropriate label or set of labels for each example.
The annotators follow established rules and criteria for their labeling task, to decide which label is the most appropriate from the available \emph{label options}.
The types of label options may vary. 
For example, annotators could be asked to make a binary choice like true or false or they may be asked for more open-ended labels such as entering any noun that most accurately describes the data.
Additionally, each example may require multiple \emph{categories} that an annotator must select a label for.
For example, in the context of a chatbot, an annotator might label whether the chatbot correctly answered a user's question and whether the response was easy for the user to understand.
In this case, there are two categories, correct response and understandable response, and each have two label options, \emph{yes} and \emph{no}, that the annotator must select from.

While performing labeling tasks, annotators often record the labels in a spreadsheet.
Figure~\ref{fig-spreadsheet} shows an example based on an actual labeling task at a large technology company.
The spreadsheet contains the data to be labeled in the left two columns (Figure~\ref{fig-spreadsheet}a), which is the transcript between a human and a chatbot.
The labels are found on the right (Figure~\ref{fig-spreadsheet}b), which consist of three different categories: relevance, quantity, and manner.
Each category has the option of being labeled \emph{y}, \emph{n}, or \emph{s} (yes, no, skip).
During the task, the annotator reads the transcript, enters the appropriate labels, and continues scrolling down the spreadsheet to label the remainder of the conversation.
Moreover, the annotator may refer to the code set, often in the form of text documents, whenever they are unsure of what label option to choose.

\subsection{Studies on Data Labeling}

Prior research has looked at data labeling and annotation tasks and focused on methods to reduce labeling effort in various ways. 
One approach is to estimate the difficulty per example to label and using that to reduce the number of examples an annotator would have to label~\cite{arora2009estimating, culotta2005AAAI}. 
Another popular approach is by developing better algorithms for selecting which examples to present to the annotator in an active learning scenario. Several prior works have surveyed such algorithms and suggested improvements~\cite{blum1997ai, fu2013survey, settles2008EMNLP, wang2011active}.
Our work differs since we focus on providing relevant information to help annotators making a labeling decision in the environment instead of quantifying labeling difficulty or choosing high-information examples to present to the labeler.

In contrast with approaches that  have looked at reducing the number of examples annotators have to label, others have focused on soliciting extra information from labelers to increase the amount of information provided with the label. 
Zaiden et al. developed a framework to extract rationales in addition to labels from annotators~\cite{zaidan2007using,zaidan2008machine}.
Others extended that approach to work in other domains~\cite{donahue2011annotator,mcdonnell2016relevant}. 
Building on \cite{zaidan2007using}, Yessenalina et al. developed a method to automatically generate rationales, thereby bypassing the need to solicit the rationales from the annotators~\cite{yessenalina2010automatically}.

An alternative approach to reducing labeling effort and soliciting extra labeling info is to determine and leverage the most consistently accurate annotators.
Donmez et al. developed a technique to select the most accurate annotators among a crowdsourced group in an effort to reduce noise in the labeled data~\cite{donmez2009kdd}. 
In a similar vein, Sheng et al. developed a method for determining which crowdsourced-labeled examples to send back to the crowd to relabel~\cite{sheng2008get}. 
Snow et al. showed that crowdsourced workers were as good as experts for natural language labeling tasks when bias correction approaches were applied~\cite{snow2008cheap}.

\subsection{Tools to Support Labeling}

As supervised machine-learning techniques continue to grow in popularity, a number of tools have emerged to connect people with data to label with annotators willing to label that data.
Tools such as Figure Eight\footnote{https://www.figure-eight.com} (previously CrowdFlower) and Hive\footnote{https://thehive.ai/} provide data labeling services to researchers and businesses.
Similarly, Mechanical Turk\footnote{https://www.mturk.com/} enables researchers to hire people to act as annotators for small tasks.
These platforms are efficient at connecting annotators to a labeling task, but they have little to no explicit support built in to support the annotator.
Our work differs because it addresses the question of what sorts of affordances do annotators need to label more efficiently and more reliably.

Researchers have also studied challenges regarding data labeling and have developed tools to supporting labeling tasks.
In the domain of chat logs, TextPrizm~\cite{Scott2017Dissertation} and Aeonium~\cite{Drouhard2017PacificVis} both provide affordances to the annotator to ease the task of labeling data. 
Aeonium provides several types of feedback to the annotator to help the annotator better assess how well their labeling conforms to the rest of the crowd.
In particular, Aeonium supports groups of annotators in identifying data that is unclear to overcome inconsistencies in labeling across the group.
Brat, a tool to support natural language processing labeling tasks, allows annotators to select and label sections of text \cite{Stenetorp2012BRAT}. 
To help annotators make consistent decisions, it uses a machine learning-based system to recommend semantic classes to the annotator, along with a probability estimate.
%

\begin{figure*}[h]
\centering
\includegraphics[scale=1.0]{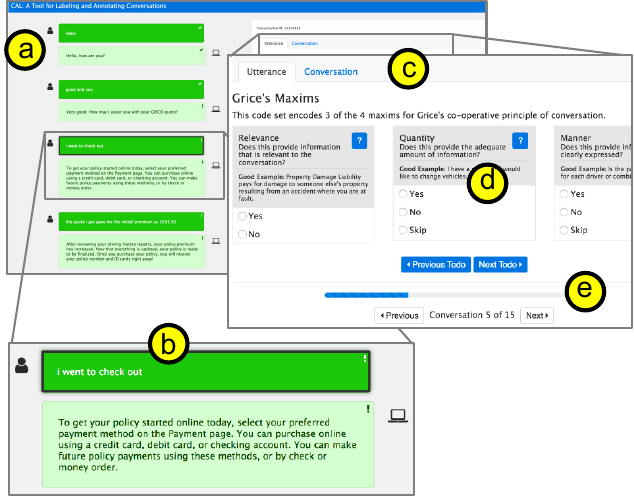}

\vspace{0pt}
\caption{%
(a) The CAL web application's data labeling view, including the (b) conversation transcript.
To label the utterances, a user selects an individual utterance and uses the (c) labeling interface to (d) select the applicable labels.
The user can track their progress with the (e) progress bar and navigate to the next/previous conversation.
}
\label{fig-cal2}
\end{figure*}

\section{Tool Design: CAL}
\label{sec-tool}

To support annotators in efficiently labeling conversational data, we designed CAL, a web application designed to enable groups of data labelers to efficiently label conversational data.
This section first details the basic features of CAL. 
Next we describe the design rationale for the novel features of CAL. 
Lastly, we describe the implementation details for CAL. 

\subsection{Basic Features}
\label{sec-tool-basic}

The essential feature of CAL is to enable an annotator to label an example from a data set.
In this paper, CAL had been configured to label conversations, so the examples to be labeled were utterances from a chat bot's logs.
First, the user clicks on the utterance that they are labeling from the conversation transcript (Figure~\ref{fig-cal2}b), which displays the labeling interface on the right (Figure~\ref{fig-cal2}c). 
Next, the user can select the appropriate labels via radio buttons, check boxes, and text boxes. 
Which label sets and label options appear for each example are defined in a configuration file supplied with the project when it is created.
The task may require them to select choices for multiple categories (e.g., Figure~\ref{fig-cal2}d shows three categories with radio buttons: Relevance, Quantity, and Manner. 
After completing their selections, they can move on to labeling the next utterance, either by clicking the Next button, using a keyboard shortcut, or by clicking the next utterance in the conversation transcript.
Utterances that have not been completely coded have a small exclamation mark, while completed ones have checkmarks.
The user continues doing this for every utterance that requires labels in the conversation.
Additionally, CAL supports annotation for the entire conversation.
If defined in the configuration file, after completing the utterance labels, the annotator will be presented with the appropriate label options for the entire conversation.
The interface for annotating conversation-level labels is similar to the utterance-level one.
The annotators continues applying labels to utterances and conversations until the progress bar (Figure~\ref{fig-cal2}e) shows 100\%, then they can click or use a keyboard shortcut to proceed to the next conversation.

\begin{figure}
\centering
\includegraphics[scale=1.0]{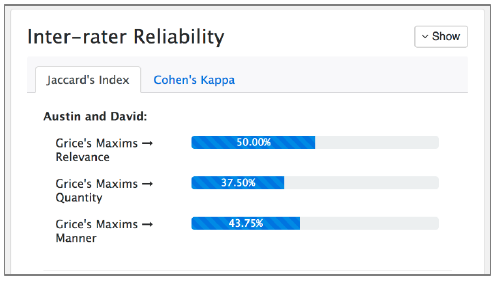}

\vspace{-3pt}
\caption{%
The inter-rater reliability view that displays the amount of agreement between data labelers. 
In this case, the two labelers have 37.5\% to 50\% agreement on three different categories using Jaccard's Index.
}
\label{fig-irr}
\end{figure}

At any time, the annotator can view the status of their labeling tasks via a project status page.
A progress bar depicts each annotator's progress towards completing the labeling tasks.
Optionally, measurements of agreement (e.g., Cohen's Kappa and Jaccard Index) can be presented, either to all annotators or to the creator of the project as shown in Figure~\ref{fig-irr}.
CAL stores where each annotator left off, so the status page provides a quick link to easily get the annotators back on task.

There are two main levels of configuration in CAL: projects and annotations.
A project configuration contains a set of data to be labeled, the annotators to do the labeling, and the assignment of code sets to utterances or conversations.
The annotation configuration defines the categories and corresponding label options, the type of label (radio buttons, check boxes, and text boxes), definitions, examples and the rules for disabling or auto selecting related labels.
If there is a flow-chart-like rule set, that is also encoded in the rule set.
Configuration is stored in a JSON format and dynamically read by the CAL application and appropriately displayed to the annotator during the labeling task.


\subsection{Design Rationale and Novel Features}
\label{sec-tool-design}

During our formative investigations in how annotators perform data labeling tasks, we identified a number of issues that annotators faced. 
First, and especially at the beginning of the annotation task, annotators often referred back to documentation on their labeling rules to better inform their decisions or the recall nuances of a particular label.
Doing so required them to repeatedly alternate between applications (e.g., a spreadsheet for their labels and a text document for the documentation), or refer to a printed rule set, which took considerable time.
Second, annotators made mistakes entering labels. 
For example, we observed annotators entering an invalid label for a category (e.g., entering ``Y'' or ``N'' when ``T'' or ``F'' is expected).
Another example, was in instances where a code set's rules included dependencies between labels (e.g., skip categories 2 and 3 if category 1's label is ``F'').
These mistakes could be caused by the annotator not being able to remember all of the labeling or simply unintentionally entering the wrong label and not realizing it.
Third, even when annotators did not make a mistake, it was time consuming and tedious for them to follow the labeling documentation in order to provide a well-informed label. 
Fourth, annotators often referred back to their previous labels, requiring them to scroll back-and-forth between their labels, and not only increasing cognitive load as they had to remember why a label was given to a particular item, but potentially also causing additional issues (e.g., inadvertently changing a label while scrolling or confusing a previous label for another).

As a result of these findings, we identified four key principles that CAL was based upon:
\begin{itemize}
\setlength\itemsep{0.0em}

\item
Provide the label ruleset and documentation in the interface.

\item
Prevent the user from selecting inapt labels.

\item 
Guide the annotator in choosing the best label based on documentation.

\item
Provide an affordance to efficiently view previous labels chosen by the annotator.

\end{itemize}

\subsubsection{Integrate labeling code set}

To eliminate the need for annotators to alternate between applications or between the computer screen and paper, CAL integrates the code set's documentation into the user interface. 
CAL displays definitions along with examples for each of the categories (see Figure~\ref{fig-cal2}d for Quantity's definition). 
This eliminates the need for annotators to refer to documentation outside the tooling.
%
%
A button to toggle this information is also provided to reduce any unnecessary clutter.
%
%

\subsubsection{Prevent inapt labels}

For some code sets, additional rules restrict which code sets, categories or even labels should be applied to a given utterance.
For example, for code sets that define label dependencies (i.e., if certain categories should be excluded due to a label option being selected, or if label options in other categories should be automatically be selected due to label option being selected), CAL provides mechanisms for enforcing those sorts of rules.
First, when selecting an utterance, CAL will only show relevant categories and label options for that utterance.
Second, after selecting a label that depends on other categories' labels, CAL will disable any inapt label options from being selected in combination with the current selection and automatically trigger other label options to automatically be selected if appropriate.
For example, selecting the ``Not Applicable'' label for one category could automatically label other categories as ``Skip.''
By enforcing these rules in the application, CAL prevents annotators from making common errors, such as selecting invalid labels.

\begin{figure}
\centering
\includegraphics[scale=1.0]{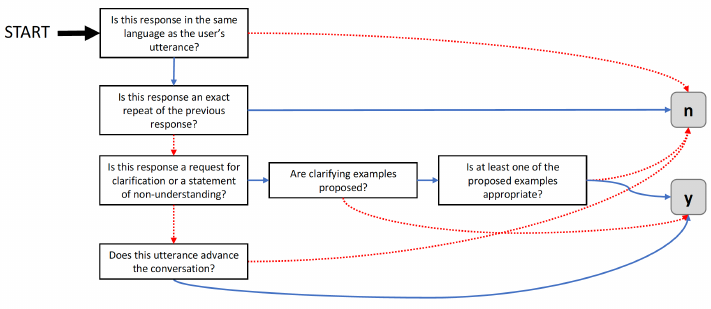}

\vspace{-4pt}
\caption{%
An example of labeling documentation that helps labelers in choosing the correct label by answering a series of yes/no questions.
The wizard feature presents these questions to the user one question at a time.
After answering all of the questions, CAL will select the label for the user and notify the user that the label was selected.
}
\label{fig-wizarddoc}
\end{figure}

\begin{figure}
\centering
\includegraphics[scale=0.9]{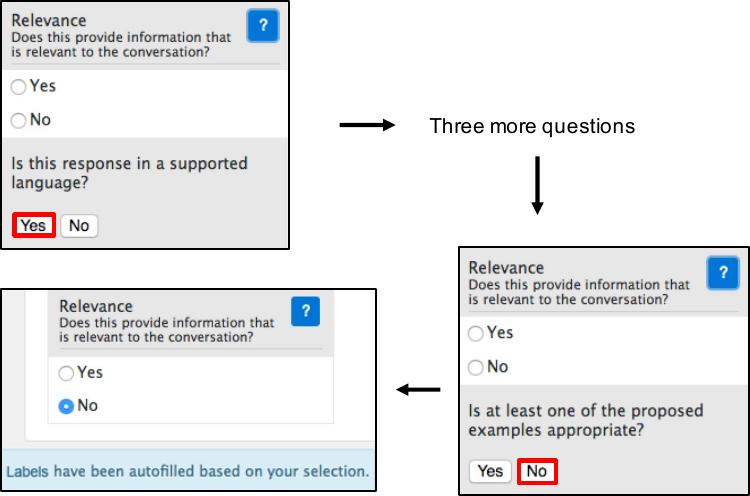}

\vspace{0pt}
\caption{%
An example of using the Wizard feature to aid an annotator in selecting the most appropriate label.
After clicking the ``?'' button to initiate the wizard, the annotator is asked a series of binary questions.
Depending on the answers provided to these questions, a label is automatically selected on behalf of the annotator.
The annotator can revisit the wizard or modify the selected label if needed.
}
\label{fig-wizard}
\end{figure}

\subsubsection{Assist in choosing the best label}

As mentioned above, to aid annotators in selecting the most appropriate label, CAL provides a wizard to guide the annotator in their selection. 
The wizard consists of a sequence of yes/no questions regarding the utterance, which after answering, will automatically select the appropriate label and display a notification that the label was automatically selected. 
This feature was inspired by documentation already used by labelers that acts as a set of heuristics by breaking down the labeling criteria into easier to understand questions. 
For example, Figure~\ref{fig-wizarddoc} shows a visualization of such a wizard from an existing project's code set's documentation (CAL does not display the visualization to users).
The questions and the outcomes must be defined in CAL when configuring the project, which allows CAL to generate the wizard.
The questions are presented one at a time in CAL's labeling interface, as depicted in Figure~\ref{fig-wizard}.

\subsubsection{Efficient viewing of previous labels}

From our formative investigations, we often observed labelers referring back to previous labels.
The labelers did this to remind themselves of what they labeled in a particular case and also in an attempt to be consistent in their labeling.
CAL provides an affordance to compare the current utterance label to a previously labeled utterance. 
To see this comparison, the labeler hovers over a potential label, which reveals a "View Previous" button. 
Clicking the button will show a recent previous utterance that this labeler had labeled with this specific label.
For example, hovering over ``Yes'' for the Relevance and clicking the ``View Previous'' button will show a comparison view of previous utterance that the labeler marked ``Yes'' in the Relevance category to the currently selected utterance.
This view is displayed below CAL's labeling interface (the otherwise unused area of Figure~\ref{fig-cal2}a in the bottom right).

\subsection{Implementation Details}
\label{sec-tool-implementation}

CAL is a written as a Node.js\footnote{http://nodejs.org/} application. 
The front-end is built using HTML 5, Embedded JavaScript\footnote{http://ejs.co/}, and Bootstrap\footnote{https://getbootstrap.com/}. 
Data is stored on a MongoDB\footnote{https://www.mongodb.com/} backend and accessed via the Flask\footnote{http://flask.pocoo.org/} microservice framework. 
For the purposes of the study, the application was running a VirtualBox\footnote{https://www.virtualbox.org/} virtual machine configured via Vagrant\footnote{https://www.vagrantup.com/}. 
To guarantee that each participant received the same treatment, we spun up a clean pre-configured Vagrant instance of the virtual machine for each participant.

\begin{table*}
\centering
\caption{%
Background and demographic information of study participants.
}
\vspace{0pt}
\includegraphics[scale=0.86]{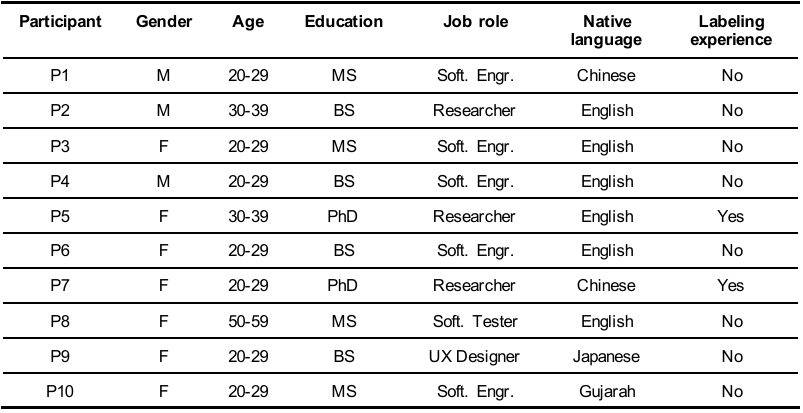}

\vspace{-6pt}
\label{tab-participants}
\end{table*}

\section{Evaluation Method}
\label{sec-method}

To address our research questions, we conducted a laboratory study of people performing data labeling tasks.
The study had a within-subjects design in which each participant experienced two treatments.
The control treatment was a standard spreadsheet using Microsoft Excel based on our prior observations in the field.
The experimental treatment was the CAL web application. To account for order effects, we blocked and balanced based on the treatment orderings.

\subsection{Participants}

Our participants were recruited by advertising on Slack groups (a corporate chat application) at a large technology company.
We did not require any particular experience or skills, but the groups we advertised to consisted mostly of researchers and software engineers.
They were provided a free lunch voucher for their participation. 
After recruiting for one week, 10 participants (7 female, 3 male) agreed to take part in the study. 
All participants had experience using Excel, but only two participants had experience with data labeling: one used Excel to label images and one used Excel to label chatbot data. 
See Table~\ref{tab-participants} for background and demographic information regarding the participants.

\subsection{Labeling Tasks}

For our study, the participants performed labeling tasks on conversations between a human and a chatbot.
These conversations consisted of real customers seeking help while attempting to purchase a product on a publicly available website.
The nature of the conversations is proprietary and cannot be disclosed.
We filtered this dataset to find recent conversations within the last 3 months and to restrict the minimum and maximum conversation length (10-20 utterances). 
After filtering, we randomly selected 30 conversations.

The participants were tasked with labeling the utterances in the conversations. 
In this context, conversations consisted of alternating utterances between the chatbot and the human. 
The labels were based on a subset of Grice's maxims~\cite{grice1975logic}, a set of established properties that are needed in an effective conversation.
For our participants, this meant labeling ``Yes'' or ``No'' for three different properties (Quality, Quantity, and Manner) regarding each utterance. 
An additional label, Topic Change, was used for human utterances that indicated whether or not the human changed conversation topics.
After labeling all of the utterances in a conversation, the participants moved on to the next conversation.

\subsection{Procedure}

Each participant took part in an individual session that lasted approximately 75 minutes.
All participants began the session by signing a consent form, listening to a summary of the study session, and filling out a background questionnaire.
They were also instructed on how to think aloud during the labeling tasks.
Next, each participant performed labeling for 25 minutes with one treatment and then performed labeling for 25 minutes with the other treatment.
Five random participants used the control spreadsheet first and five participants used the CAL application first.
The participants were given a training session on the labeling task with explanations on the labeling criteria.
They were also provided documentation on the labels (Grice's maxims~\cite{grice1975logic} and topic change) consisting of definitions and examples in the form of PDFs that they could refer back to at any time.
They were given a demonstration on each tool before using it.
%
After completing each 25 minute task, they were a given a cognitive load questionnaire regarding the task and a usability questionnaire regarding the tool they used.
At the end of the session, all participants took part in a semi-structured interview to discuss benefits, challenges, and possible improvements for both CAL and the spreadsheet.

\subsection{Data Collection}

During the study, the computer screen and audio of the participants was recorded.
The CAL application also logged the participant's interactions (e.g., label selections and button clicks).
For the cognitive load questionnaire, we used a well-validated short shelf-report instrument based on Cognitive Load Theory~\cite{Paas2003}. 
The instrument measures a person's cognitive load during a task with a 7-point Likert scale and has been shown to correlate with complex measurements, such as NASA TLX~\cite{Windell2007AERA}, and with physiological measurements, such as heart rate~\cite{Paas1994PMS}.
For the usability questionnaire, we used the widely used system usability scale (SUS)~\cite{Brooke1996}.
This questionnaire consists of 10 Likert-scale questions that produce a single score that can be used to compare systems in terms of usability.

\section{Results}
\label{sec-results}

The 10 participants completed two sequences of labeling, each using CAL and the spreadsheet.
Collectively, the participants made 839 labels for utterances in 30 different conversations.
The remainder of this section describes the results for our 4 research questions.

\begin{figure}
\centering
\includegraphics[scale=1.0]{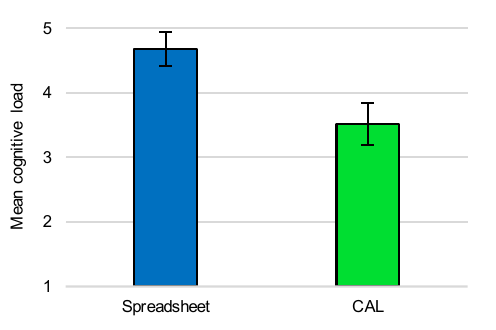}

\vspace{-10pt}
\caption{%
CAL users reported significantly lower cognitive load than spreadsheet users (lower is better).
Whiskers denote standard error.
}
\label{fig-results-cognitive}
\end{figure}

\begin{figure}
\centering
\includegraphics[scale=1.0]{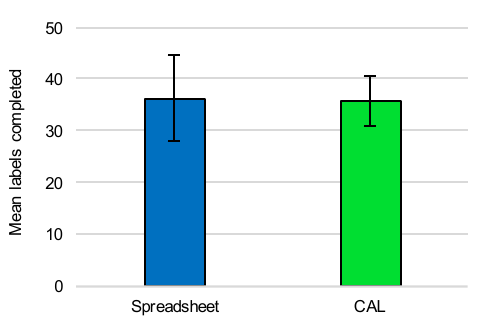}

\vspace{-10pt}
\caption{%
Overall, spreadsheet and CAL users did not exhibit a significant different in the number of labels completed.
Whiskers denote standard error,
}
\label{fig-results-completion}
\end{figure}

\subsection{RQ1 Results: Cognitive Load}

As Figure\ref{fig-results-cognitive} shows, participants using CAL reported considerably lower cognitive load than spreadsheet users.
Indeed, every individual participant reported a lower or equal cognitive load using CAL than when they used the spreadsheet.
The results of a Mann-Whitney $U$ test showed that CAL users reported a significantly lower cognitive load than spreadsheet users ($U = 19$, $Z = 2.3$, $p < 0.05$).

\subsection{RQ2 Results: Task Progress}

As Figure~\ref{fig-results-completion} shows, CAL users labeled a similar number of utterances as spreadsheet users.
In fact, there is less than a 2\% difference in the average number of labels completed between the treatments, despite all participants having experience with Excel and no experience with CAL.
The number of labels completed did not differ significantly between CAL and spreadsheet users.
Each individual participant did complete more labels during their second task than their first task, regardless of the treatments, which suggests a learning effect (see Section~\ref{sec-method-limitations} for a discussion of the study limitations).

\subsection{RQ3 Results: Usability}

\begin{figure}
\centering
\includegraphics[scale=1.0]{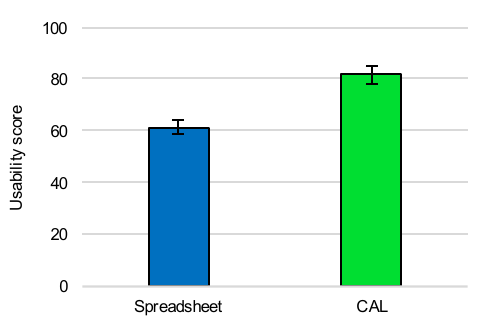}

\vspace{-8pt}
\caption{%
Participants rated CAL to be significantly more usable than the spreadsheet on the SUS questionnaire~\cite{Brooke1996} (higher is better). 
Whiskers denote standard error.
}
\label{fig-results-usability}
\end{figure}

As Figure~\ref{fig-results-usability} shows, CAL users rated the usability of CAL considerably higher than spreadsheet users on the SUS questionnaire~\cite{Brooke1996}, which consists of 10 Likert-scale questions regarding the tool's effectiveness, efficiency, and satisfaction.
In fact, all 10 participants rated CAL higher than they rated the spreadsheet in each of the questions on the SUS questionnaire.
The results of a Mann-Whitney U test shows that CAL users reported a significantly higher usability score than spreadsheet users ($U = 5.5, Z = 3.3, p < 0.001$).

\begin{figure}
\centering
\includegraphics[scale=1.0]{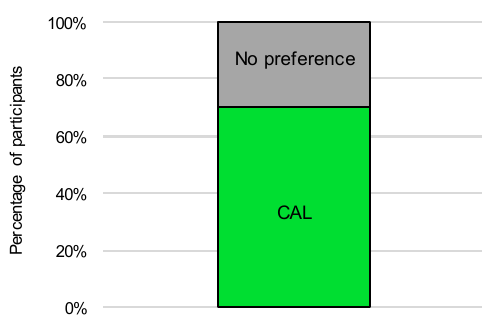}

\vspace{-8pt}
\caption{%
When asked which tool did the participant prefer using, the majority chose CAL.
No participant chose the spreadsheet.
}
\label{fig-results-preference}
\end{figure}

\subsection{RQ4 Results: Preference}

As Figure~\ref{fig-results-preference} shows, the majority of the participants stated they would prefer using CAL over the spreadsheet in the future.
Of the 10 participants, 7 participants said they would prefer to use CAL and 3 participants said ``it depends'' on the context.
No participant preferred using the spreadsheet over CAL.
The details of their preferences are explained in Section~\ref{sec-discussion}.

\section{Discussion}
\label{sec-discussion}

Overall, the quantitative results from our user evaluation of CAL were highly favorable. CAL significantly reduced users cognitive load during the labeling tasks. 
Furthermore, CAL users took roughly the same amount of time on tasks as spreadsheet users.
After the labeling tasks, users rated CAL to be significantly easier to use based on the SUS questionnaire and when asked which tool they preferred, the majority of users chose CAL.

The remainder of this section will report our analysis of the qualitative evidence from the think aloud data and interview responses in an effort to better understand our results.

\subsection{CAL eliminates typing}

The CAL interface eliminated the need for typing the specified label.
P2 and P6 both expressed their satisfaction with how CAL eliminated the need to type on the keyboard.
P2 continued by indicating it was beneficial to not have to continuously switch between keyboard and mouse, like he had to do with the spreadsheet.
P3 had a similar thought, saying that he could no longer make a mistake while typing because CAL didn't require any typing.

However, the reliance on using the mouse did have performance implications.
P10 said she believed that using CAL took more clicks than the spreadsheet did.
P5 and P7 responded to interview questions by indicating that the spreadsheet is faster to input the labels into.
P5 continued by saying that she felt faster using the spreadsheet but would prefer CAL if there were keyboard shortcuts (although shortcuts were implemented, our tutorial on CAL did not focus on them).

\subsection{CAL provides assistance for labeling}

CAL users overwhelmingly found the wizard feature to be beneficial.
In fact, participants 6 of the participants made verbal remarks about the usefulness of the feature, including:

\begin{ParticipantQuote}

\item[P2:]
`` I like the thing with the question mark where you can decide whether it is relevant or not. With the spreadsheet, you have to go look at the documentation.''

\item[P3:]
``I liked that I got the help when I needed to by clicking on those question marks if I was ever stuck. It was a lot easier than going back to the documentation.''

\item[P4:]
``Having the integrated help so that you can step through the questions if you're unsure about something, it helps you get there.''

\item[P8:]
``I liked that if I didn't know what to pick, it would take me through a series of questions and pick the answer for me.''

\end{ParticipantQuote}

The other labeling features provided by CAL were also found to be useful for participants.
The autofilling of labels was deemed helpful by both P7 and P8.
Although the View Previous feature was used sparingly by our participants, P5 expressed multiple positive sentiments:

\begin{ParticipantQuote}

\item[P5:]
`` I think this View Previous is really good... I think it is a really good idea. Even when I filled out the first one, I wanted to view my previous labels... I think that is super valuable.''

\end{ParticipantQuote}

\noindent
On the other hand, P7 was observed manually going back to view previous labels without using this feature, which indicates future work to make this feature easier to use.

\subsection{Spreadsheet was not user friendly}

Participants complained about various ways in which the spreadsheet was difficult to use.
Most commonly, they complained about how difficult the spreadsheet was to read.
In fact, 5 participants verbally expressed this sentiment throughout their session, such as:

\begin{ParticipantQuote}

\item[P1:]
``It is difficult for me to read.''

\item[P6:]
``The cells do bombard me, then I have to remember which column is going for which one.''

\item[P10:]
``I think the spreadsheet hurts your eyes after a while. I think it is because it is not user friendly.''

\end{ParticipantQuote}

\noindent
P4 and P5 drew comparisons to their experience with using CAL:

\begin{ParticipantQuote}

\item[P4:]
``Being able to see it laid out in text conversation rather than having to read cells, that made it easier.''

\item[P5:]
``This one is like ahhh because I don't have to strain my eyes [like in the spreadsheet].''

\end{ParticipantQuote}

\subsection{Spreadsheet showed more information}

The spreadsheet shows more information on the screen at any given time than CAL.
Based on the participants' feedback, one of the biggest advantages that the spreadsheet had over CAL is that it allows the user to quickly scan the labels for multiple utterances without any actions.
This was summarized well by P8:

\begin{ParticipantQuote}

\item[P6:]
``For the spreadsheet, it was easier to compare all my responses overall.''

\end{ParticipantQuote}

\noindent
During the interviews, 5 of the participants suggested that CAL could be improved by showing the label for the utterance and the response simultaneously.  
P3 also requested that items in CAL be closer together to reduce the amount of scrolling needed.

\subsection{Limitations}
\label{sec-method-limitations}

Our study had a number of limitations to generalizability that are common to laboratory studies.
First, the tasks may not generalize to other situations of labeling data, but we did base the tasks on actual labeling tasks being conducted for industrial use and we used data from an actual product.
Second, it is not well understood how our sample of participants represents our target population.
To partially mitigate this, we recruited participants from a large technology company.
Third, using spreadsheets may not be representative of how researchers do conversation labeling tasks, but our formative investigations found it to be the most used tool.
Fourth, the participants may have concluded that CAL was our design and thus biased their behavior and responses.
Fifth, since this was a within-subjects study, there may be order effects of the treatments and tasks; however, we counterbalanced the treatment order to control for this effect.

\section{Conclusion}
\label{sec-conclusion}

In this paper, we introduced the novel CAL tool to support annotators in labeling data more efficiently.
We designed the tool based on 4 key design goals and implemented a production-quality web application.
An evaluation study comparing our CAL tool to a standard spreadsheet made the following key findings:
\begin{itemize}
\setlength\itemsep{0.0em}

\item
RQ1: CAL users reported significantly lower cognitive load during labeling tasks.

\item
RQ2: CAL had no noticeable effect on the time taken to label data.

\item 
RQ3: Users rated CAL to be significantly more usable than the spreadsheet.

\item
RQ4: Most users preferred using CAL, while two reported ``it depends'' (none preferred the spreadsheet).

\end{itemize}

CAL aims to make considerable advances in supporting human annotators in the tedious task of manually labeling data to be used in machine learning-based systems.
In the future, CAL could utilize machine learning to provide suggestions of which label to select based on the annotator's recent labels.
Such a technique could also be applied to CAL's \emph{View Previous} feature, by using an intelligent recommendation system to display relevant data that the annotator recently labeled.
Additionally, our study revealed that annotators begin showing signs of fatigue after an hour of annotating, which suggests that CAL should monitor for behavior that is indicative of fatigue, such that it can ensure the labels are accurate.
These future works have tremendous potential to transform the way in which human annotators label data.
%
%

\bibliographystyle{ACM-Reference-Format}
\bibliography{refs}
\end{document}